\documentclass[sigconf]{acmart}

\copyrightyear{2024}
\acmYear{2024}
\setcopyright{rightsretained}
\acmConference[CHI EA '24]{Extended Abstracts of the CHI Conference on Human Factors in Computing Systems}{May 11--16, 2024}{Honolulu, HI, USA}
\acmBooktitle{Extended Abstracts of the CHI Conference on Human Factors in Computing Systems (CHI EA '24), May 11--16, 2024, Honolulu, HI, USA}
\acmDOI{10.1145/3613905.3650914}
\acmISBN{979-8-4007-0331-7/24/05}

\usepackage{subcaption}

\begin{document}

\title{Mitigating Ageism through Virtual Reality: Intergenerational Collaborative Escape Room Design
}

\author{Ruotong Zou}

\affiliation{%
  \institution{Southern University of Science and Technology}
  \city{ShenZhen}
  \country{China}
}
\email{zourt2020@mail.sustech.edu.cn}

\author{Shuyu Yin}
\affiliation{%
  \institution{The Hong Kong Polytechnic University}
  \city{Hong Kong}
  \country{China}}
\email{shuyu1.yin@polyu.edu.hk}

\author{Tianqi Song}
\affiliation{%
  \institution{National University of Singapore}
  \city{Singapore}
  \country{Singapore}
}
\email{tianqi_song@u.nus.edu}

\author{Peinuan Qin}
\affiliation{%
 \institution{National University of Singapore}
 \city{Singapore}
 \country{Singapore}}
 \email{e1322754@u.nus.edu}

\author{Yi-Chieh Lee}
\affiliation{%
  \institution{National University of Singapore}
  \city{Singapore}
  \country{Singapore}}
  \email{yclee@nus.edu.sg}




\renewcommand{\shortauthors}{Trovato and Tobin, et al.}


\begin{abstract}
As virtual reality (VR) becomes more popular for intergenerational collaboration, there is still a significant gap in research regarding understanding the potential for reducing ageism. Our study aims to address this gap by analyzing ageism levels before and after VR escape room collaborative experiences. We recruited 28 participants to collaborate with an older player in a challenging VR escape room game. To ensure consistent and reliable performance data of older players, our experimenters simulated older participants following specific guidelines. After completing the game, we found a significant reduction in ageism among younger participants. Furthermore, we introduce a new game mechanism that encourages intergenerational collaboration. Our research highlights the potential of VR collaborative games as a practical tool for mitigating ageism. It provides valuable insights for designing immersive VR experiences that foster enhanced intergenerational collaboration.

\end{abstract}

\maketitle

\section{Introduction}
Age-based prejudice, commonly called ageism, encompasses stereotyping, prejudice, and discrimination directed against individuals based on their age \cite{burnes2019interventions}. Ageism has been found to exert a detrimental impact on health \cite{chang2020global}. The prejudice even extends to professions dedicated to assisting older individuals, exerting a significant negative influence on the physical and mental well-being of their older patients \cite{nelson2010ageism}.
Furthermore, ageism fosters negative perceptions of the aging process, contributing to feelings of anxiety and avoidance towards the older \cite{levy2019ageing}. Therefore, addressing the reduction of ageism is of great significance, especially in the context of the global aging trend \cite{world2016multisectoral}. 

Intergenerational collaborative projects have consistently demonstrated their positive impact on ageism mitigation \cite{springate2008intergenerational}. For example, intergenerational educational programs that combine educational courses with social intergenerational interaction have shown promise in reducing ageism \cite{murphy1986changing, rubin2015, loe2013digital}. In addition, intergenerational volunteer programs such as doing housework for the older can also reduce ageism \cite{dooley1990improving}. However, these programs have encountered issues related to health, transportation, weather, or other restrictions for safe social connection, especially during the COVID-19 pandemic \cite{jarrott2022intergenerational}. Intergenerational programs are routinely suspended or delayed not only during pandemics but also in adverse weather conditions or when infectious diseases become a concern \cite{united2018all}. To address these limitations, we introduce virtual reality (VR) as a potential solution.

Virtual reality refers to an artificial environment that individuals experience through sensory stimuli \cite{jerald2015vr}, providing them with the sensation of being in a real place \cite{sherman2018understanding}. By implementing intergenerational interventions within VR environments, the older can participate in experiments at their own residences with comfort and safety \cite{xu2023designing}. 
Immersive VR has been increasingly used to improve the well-being of older adults in recent years \cite{van2021virtual}. It is applied to address a variety of issues related to aging, such as aged care \cite{baker2020evaluating, saredakis2020using, waycott2022role}, social interaction \cite{xu2023designing}, and cognitive training \cite{coyle2015computerized, hill2017computerized}. Nevertheless, there exists a notable gap in the literature concerning the application of VR to mitigate ageism.

While a limited number of previous studies have explored how ageism can be mitigated through virtual embodiment \cite{banakou2018virtually, ayalon2023combatting} or perspective-taking \cite{OH2016398}, there is still a gap between examining the effectiveness of intergenerational collaborative interventions in reducing ageism within VR environments. To address the gap, our research endeavors to harness VR to develop an intergenerational collaborative game specifically designed to combat ageism.

Current VR intergenerational collaborative intervention faces the challenge of social dominance that may result in frustration and worse collaboration experience \cite{xu2023designing}. In response to this challenge, we have designed a game mechanism to reduce social dominance and improve collaboration. This mechanism is based on three distinct modules, each designed to target different types of abilities: fluid cognitive ability, crystallized cognitive ability \cite{cattell1963theory}, and physical activity ability such as reaction and movement speed \cite{hodgkins1962influence}. By utilizing the respective strengths of players from different age groups, our game reduces social dominance and encourages effective intergenerational collaboration.

This research makes contributions to the area of collaborative games based on virtual reality to diminish ageism. We conducted a comprehensive examination of how a collaborative VR escape room game can effectively reduce ageism and alter young people's perspectives of the older using a mixed-method experiment. Additionally, we present a new game mechanism with cognitive and physical activity modules designed to promote intergenerational collaboration. We put forward the following research questions: \textbf{RQ1-a. Will the VR intergenerational collaboration procedure have an effect on the younger participants' levels of ageism?} \textbf{RQ1-b. How does the collaborative procedure influence factors of ageism, particularly with regard to aspects of threat, warmth, and competence?} \textbf{RQ2. How does the game design influence collaboration and social dominance?}

\section{Related Work}
In this section, we identify three factors of ageism according to two ageism models. Then, we present literature on intergenerational interventions for ageism reduction and VR for intergenerational collaboration. Finally, we introduce the cognitive and physical activity theory of our collaboration strategy. 

\subsection{Factors of Ageism}

Integrated Threat Theory (ITT) suggests that fear and danger are linked to prejudice, including ageism. ITT identifies four sources of threat that can lead to prejudice \cite{stephan2013integrated}, and states that
older people are often discriminated against due to perceived threats in terms of resource competition, conflicting values, and intergroup anxiety when interacting with younger people. Additionally, the Stereotype Content Model (SCM) proposes two key dimensions that shape perceptions \cite{fiske2018model}: warmth and competence, which can be divided into four social groups: (1) high in both – pride-inducing, (2) competent but not warm – envied, (3) warm but incompetent – pitied, (4) low in both – contemptible. older individuals are often seen as warm yet incompetent, leading to societal pity stereotypes. Drawing from these theoretical foundations, our study identifies threat, warmth, and competence as key factors contributing to ageism.

\subsection{Intergenerational Intervention for Ageism Reduction}
Contact theory, as originally proposed by Allport \cite{allport1954nature}, posits that high-quality intergroup interaction can lead to a reduction in prejudice.
In the context of ageism reduction, collaboration is a key component of intergroup cooperation. 
Multiple intergenerational studies have shown a significant positive effect on reducing ageism. For example, libraries have successfully used techno-creative activities to combat age-based stereotypes \cite{romero2017intergenerational}. Additionally, collaborative alternate reality games are effective in reducing ageism \cite{hausknecht2017blurring}. Nevertheless, these collaborative projects are confronted with issues such as health, weather, and transportation, which greatly impede the continuity of these initiatives \cite{tassinari2022use}. VR technology can address these problems by providing a virtual immersive environment that allows for intergenerational collaboration without the restrictions of physical limitations, thus ensuring the safety and comfort of the users.



\subsection{VR for Intergenerational Collaboration}
Previous research has consistently demonstrated the potential of VR to improve mutual understanding between individuals of different ages. For instance, in activities such as intergenerational learning \cite{su14106067} and buddy biking \cite{hoeg2023buddy}, both older and younger participants have reported enjoying the social aspects and gaining a better understanding of one another. Baker et al. \cite{baker2019interrogating} also show how VR can be used to adapt body movements, effectively concealing physical disabilities such as Parkinson's disease, thus allowing older users to present themselves as they wish in VR communication contexts. Wei et al. \cite{wei2023bridging} also conduct research into enhancing intergenerational communication, offering valuable design implications for more effective VR interactions between grandparents and grandchildren. Additionally, several studies dedicate efforts to the design of VR intergenerational games aimed at engaging individuals of different age groups \cite{chou2022empirical, khoo2008age, khoo2010designing}. 
However, there is still a lack of research on task design within VR intergenerational games, as well as the effect of VR intergenerational collaboration on ageism reduction.

\subsection{Cognitive and Physical Activity Theories of Collaboration Strategy}

The phenomenon of social dominance in collaborative projects can lead to potential challenges such as reduced enthusiasm and engagement \cite{xu2023designing}. To address this issue and ensure a balanced level of challenge within the game, we propose a novel game mechanism that divides the game into three parts: crystallized ability assessment module, fluid ability assessment module, and physical activity module. Crystallized ability is the skill set and knowledge base that a person has accumulated over time \cite{cattell1963theory}. It tends to increase with age, reaching its peak around the age of 60 and then gradually decreasing \cite{salthouse2012consequences}. As a result, older adults have relatively balanced or even stronger crystallized abilities than younger adults. Fluid ability, on the other hand, is about reasoning, memory, and reaction speed \cite{cattell1963theory}. The ability shows a monotonic decrease with the increase of age from 20 years old, so younger individuals are usually better at this \cite{salthouse2012consequences}. Also, for physical movement, the reaction time \cite{hultsch2002variability} and movement time \cite{hodgkins1962influence, pierson1958movement} increase with age.


\section{Methods}

\begin{figure*}
  \centering
  \begin{subfigure}{0.45\textwidth}
    \centering
    \includegraphics[width=\linewidth]{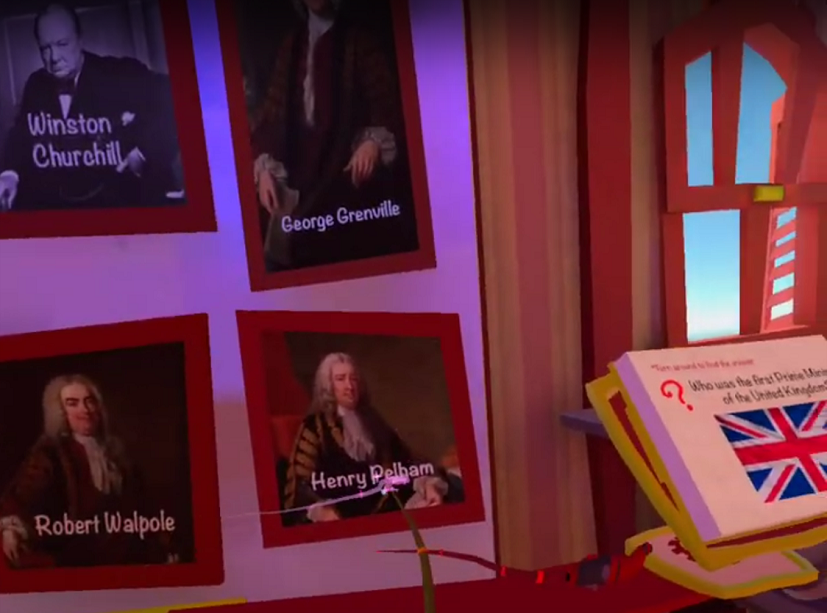}
    \caption{Quiz Task}
    \label{fig:image1}
  \end{subfigure}
  \begin{subfigure}{0.45\textwidth}
    \centering
    \includegraphics[width=\linewidth]{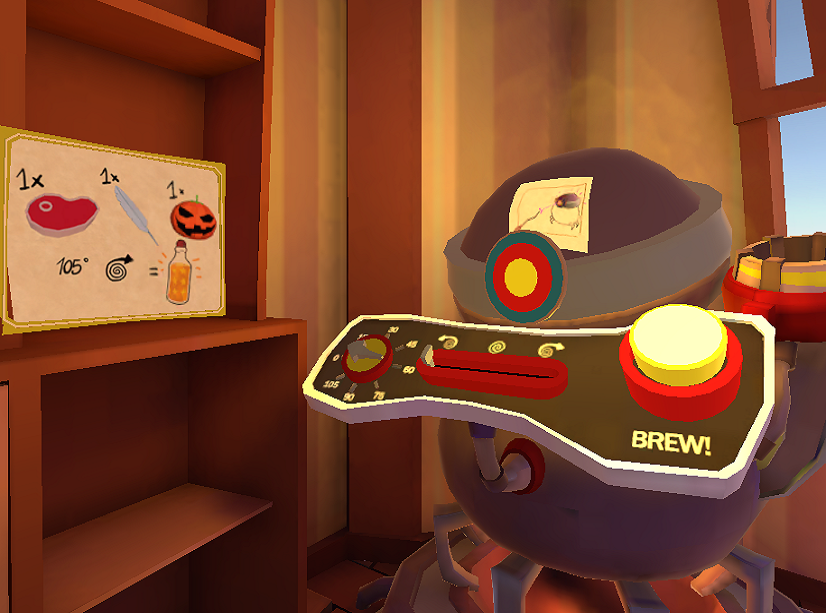}
    \caption{Cooking Task}
    \label{fig:image2}
  \end{subfigure}
  
  \vspace{1em} 
  
  \begin{subfigure}{0.45\textwidth}
    \centering
    \includegraphics[width=\linewidth]{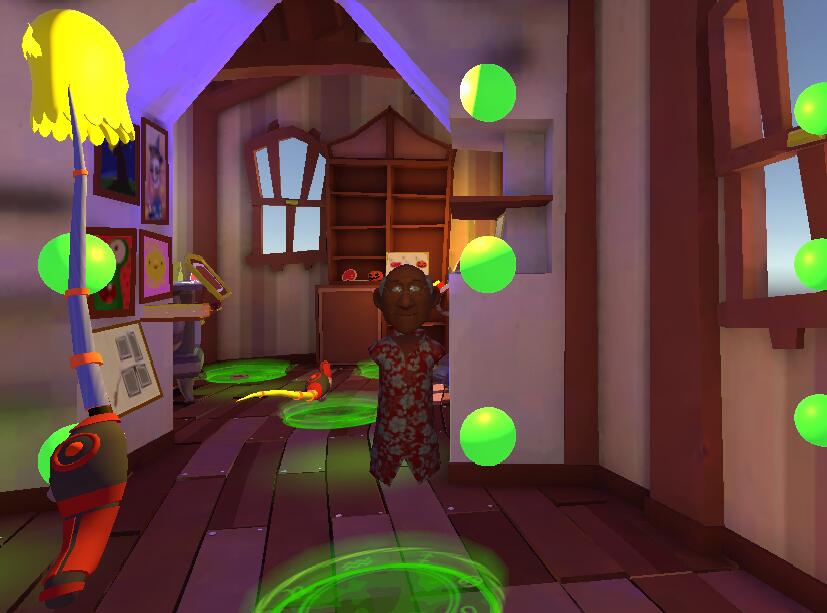}
    \caption{Shooting Task}
    \label{fig:image3}
  \end{subfigure}
  \begin{subfigure}{0.45\textwidth}
    \centering
    \includegraphics[width=\linewidth]{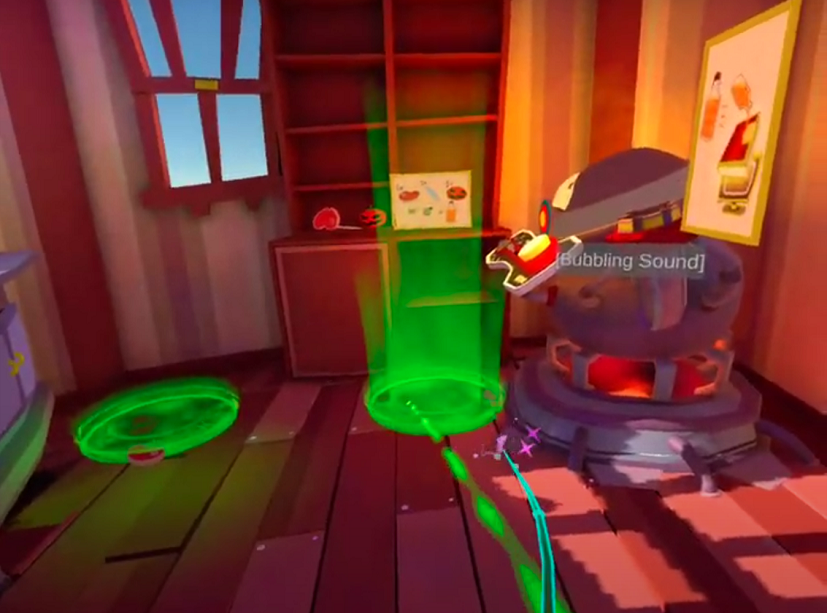}
    \caption{Teleportation}
    \label{fig:image4}
  \end{subfigure}
  \caption{Screenshots of Tasks in VR Escape Room Game from Players' Perspective}
  \label{fig:2x2layout}
\end{figure*}

Based on our research questions and literature review, we developed a multiplayer VR escape room game for intergenerational collaboration. Each team comprises a young and an older player who must work together to complete tasks and successfully escape from the virtual room. 
The participants were informed that they would collaborate with an older participant. Our experimenters who acted as older individuals adopted the older avatar in the game and utilized a voice changer to convincingly embody the characteristics of an older person.

\subsection{VR Task Design}


The VR collaborative game comprises three modules.
The crystallized ability assessment module includes four quiz tasks to evaluate participants' accumulated knowledge. As depicted in Figure \ref{fig:image1}, the quiz task requires players to use their magic wands to touch the correct answers. The questions cover topics related to history, health, and daily life. The fluid ability assessment module, including a cooking task and several smaller tasks, requires participants to solve problems by searching for clues in the room. In the cooking task, players are tasked with locating a recipe within the virtual environment and meticulously following the instructions to concoct a magic potion, as demonstrated in Figure \ref{fig:image2}. This involves adding the correct ingredients to the cauldron and adjusting associated parameters. Finally, in the physical activity module, the shooting task presents players with the challenge of using their magic wands to target and shoot ghost models that appear at regular intervals of every 4 seconds. The models spawn randomly at one of the green balls' positions, as shown in Figure \ref{fig:image3}. Upon completing a task, players can push the stick on the handler forward to teleport to the next task site, as demonstrated in Figure \ref{fig:image4}. 

The game mechanism is beneficial for reducing social dominance. In the crystallized ability assessment module, we intentionally design tasks that provide a slightly higher level of challenge for younger participants. Conversely, the fluid ability assessment module and physical activity module naturally align with the strengths of younger individuals. The mechanism leverages the unique strengths of participants from distinct age groups, fostering a more inclusive and equitable collaborative experience.




The game was developed on the Unity platform, and we utilized the XR Interaction Toolkit to enable interactive elements and enhance the user experience. Our target platform for the game is the Meta Quest 2 headset, which offers a per-eye resolution of 1832x1920 and a refresh rate of 90 Hz.

\subsection{Performance Regulations}
The experimenter, acting as an older participant, performed steadily during the game process. The performer correctly answered all quiz questions within the crystallized ability assessment module. Within the fluid ability assessment module, the performer gave a slightly slower response and was responsible for figuring out the adjustment of the cauldron. Also, through the statistics data in previous research \cite{hodgkins1962influence, hultsch2002variability}, the average reaction and movement time of older adults are about 1.3 times longer than those of the younger. In the shooting game, our younger experimenters typically responded within about 2 seconds. As a result, the performer of older individuals took about three seconds to shoot the ghost. Although experimenters were familiar with the game before the experiments, when acting as the older, they intentionally disregarded the familiarity. Instead, they rethought the problems and pretended to play the game for the first time. This trick avoided unnatural responses and improved the credibility of performance.

\subsection{Participants}
A total of twenty-eight participants were recruited from the college and successfully completed the experiment. Each participant was instructed to collaborate with an older participant in the VR escape room game. Considering the older performer's proficiency in the Chinese language, we established the following inclusion criteria: (a) participants aged between eighteen and thirty years old and (b) possessing the ability to communicate fluently in Chinese. The participants' average age was 22.96 years old (Standard Deviation = 1.37). All participants were college students, comprising ten males (35.7$\%$) and eighteen females (64.3$\%$). The study received prior approval from the university's IRB before commencing the experiment.

\subsection{Procedure}
Before reaching the research location, participants were instructed to perform the Implicit Association Test (IAT) and complete the online pre-experiment questionnaire. Upon their arrival for the gaming experiment, the overseeing researcher offered assistance and instructed them to review the guidelines for controlling the VR handlers. Following this, the participants entered the tutorial scene in IVE to acquaint themselves with the handler manipulations and view their avatars as well as their partners' ones. Subsequently, they moved on to the main game scene, where they were required to complete various tasks and escape from the virtual room. During the experiment, participants 
were allowed to have a rest whenever they felt uncomfortable. After completing the game, participants were directed to take the IAT test on a computer, fill out the post-experiment questionnaire, and participate in an interview.

For the VR game experiment, we had two experimenters involved. One experimenter acted as the older player, while the other was responsible for guiding the participants, ensuring their safety, and closely monitoring the game process throughout the experiment. The user experience within the game was observed and recorded using the Unity platform.

\subsection{Measurement}

The quantification of ageism is about evaluating three distinct dependent variables: threat, warmth, and competence. These variables collectively contribute to the ageism levels. Specifically, when individuals experience enhanced perceptions of warmth, competence, and a diminished sense of threat concerning the older, the level of ageism decreases. To explore participants' pre-experiment and post-experiment perceptions of the older participants' threat, competence, and warmth, we requested their feedback using a 7-point Likert scale. 
Drawing from the Stereotype Content Model \cite{fiske1999dis}, the warmth dimension is assessed via four traits: sincere, good-natured, warm, and tolerant. 
In the context of competence, the traits considered are intelligence, confidence, competition, independence, and competence. 
The perception of threat can be assessed via four traits: uneasy, tense, anxious, and afraid \cite{stephan1999anxiety, tausch2007individual}. As a result, we put these traits into our questionnaire and asked questions mirroring the structure, \textbf{"As perceived by society, how sincere are individuals aged over 65?"}. Moreover, to evaluate participants' explicit attitudes regarding ageism, we selected eight negative-positive question pairs from different question clusters based on Kogan's prior research \cite{kogan1961attitudes}. 

\subsubsection{Age Implicit Association Test}
The Implicit Association Test (IAT) is a tool used to measure people's automatic associations between concepts and attributes, which can uncover their implicit social cognition \cite{greenwald1998measuring}. In this study, participants were asked to quickly classify the faces of young or old people with positive or negative words. This process, known as the Age IAT, was used to measure participants' implicit levels of ageism by analyzing the speed and accuracy of categorizing different age groups with various attributes. 
In this study, the sequence of the combined tasks was randomly assigned to participants to reduce their effects on test results \cite{nosek2005understanding}. 
The Overall D-score was used for measurement in our experiment. The Age IAT used in this study was taken from the Millisecond Test Library.

\subsubsection{Interview}
The interviews were semi-structured and concentrated on three key aspects: 1) Measuring the difficulty of crystallized and fluid ability assessment modules. 2) Exploring social dominance and collaboration among participants. 3) Exploring the younger participants' change in stereotypes towards the older.

At the start of the interview, participants were asked to share their views on the challenging level of each task and the overall game and to provide the reasoning behind their perceptions. Furthermore, to assess the quality of collaboration, participants were requested to rate their collaborative experiences with their respective partners and elucidate the reasons for their evaluations. Subsequently, we asked
whether participants felt they had played a dominant role in the collaboration and their reasons. Moreover, we inquired whether participants found the collaboration and their partner exceeded their expectations and how these experiences influenced their attitudes toward the older.


\section{Results}

\subsection{Ageism Levels (RQ1-a)}

As demonstrated in Figure \ref{fig:skeptical}, a significant reduction in ageism levels for participants' explicit attitudes was also evident (t = -6.40, p < 0.001). The Aged IAT Test displayed a significant decrease as well (t = -2.13, p < 0.05). These results collectively indicate that the collaborative experience had a significant impact on both explicit and implicit attitudes towards reducing ageism. 

\subsection{Attitudes towards the VR older Player (RQ1-b)}

\begin{figure}
\includegraphics[width=3in]{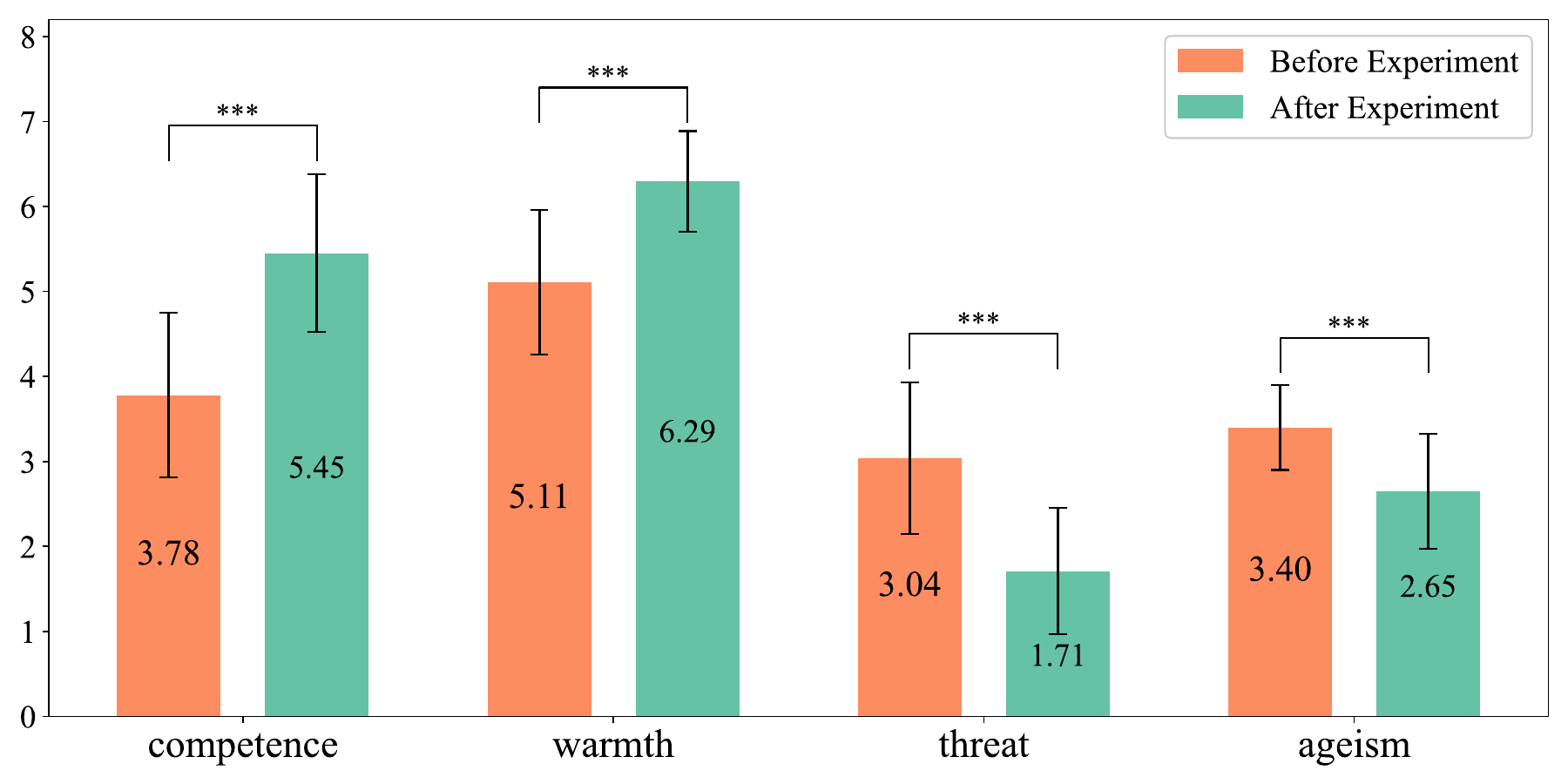}
\caption{Participants' Explicit Attitude Towards the older Before and After Experiment}
\label{fig:skeptical}
\end{figure}

We noted a significant increase in competence levels (t = 6.00, p < 0.001, Figure \ref{fig:skeptical}), warmth levels (t = 5.94, p < 0.001), and a noteworthy decrease in threat levels (t = -5.99, p < 0.001). These findings underscore the substantial impact of collaborative participation within the VR escape room game on the three crucial variables associated with ageism, ultimately resulting in a significant reduction of ageist attitudes. 


\subsection{Qualitative Analysis}
\subsubsection{Game Difficulty and Collaboration Experience (RQ2)}

The perceptions of the difficulty of the game module by the participants aligned with our hypothesis. Eighteen participants considered the crystallized ability assessment module to be more demanding for younger participants and relied on their older partners in this module. As expressed by P8, "\textit{I think they (the older) are closer to the time period in the history question, so they know more about the past. In terms of health, they should be more knowledgeable because they are older, right? I relied on him in this task."} 
Sixteen other participants shared similar opinions, suggesting that older individuals are more acquainted with history and possess greater health knowledge. Conversely, in the fluid ability assessment module and physical activity module, 23 participants believed that the older players faced more challenges and provided assistance because they had quicker responses to finding clues and reasoning. For example, P11 commented, \textit{"In the shooting task, he did not react quickly when the ghost toy spawned sometimes. I reminded him when this happened."} 

Regarding overall game difficulty, most participants perceived that the game presented slightly greater challenges for older people. The primary reason is that the game initially needs fluid ability to adapt to the virtual environment, which is typically more aligned with the skills of younger players. Participant P28 articulated, \textit{"The older usually do not have much contact with this kind of game, digital games, so they may take long to react in the experiment."} Despite the slight difficulty inbalance, eighteen participants mentioned that they acted as a collaborator in the game rather than a dominator, which accorded with the observation of the performer of the older player. These younger players believed that they were equal with their older partners, each bringing their unique strengths to the collaboration. As P25 said, \textit{"I think we worked together mostly in the game. He did well in the quiz. I think I was collaborating with him instead of helping him".} 
These insights underscore the reduction of social dominance observed in the experiment.

Participants were also asked to rate the collaboration with their partner from 1 - 7 (1 for extremely bad, 7 for extremely good). All participants rated 5 - 7 (M = 6.17, SD = 0.72), and 26 participants stated that they would have worse game performance without collaboration because they supported each other in the game. 
P6 stated, \textit{"He impressed me in the quiz task. He was knowledgeable. I could not have passed the quiz so smoothly without him. I feel that we established effective communication in the game."}  
These participants' views underscore the game's high-quality collaboration and its significant role in enhancing the gameplay experience.
These findings indicate the effectiveness of our design in leveraging the strength of players in distinct age groups, further reducing social dominance and fostering high-quality collaboration.

\subsubsection{Challenging Social Stereotypes about older}
The game effectively challenged two types of social stereotypes. Ten participants changed their attitudes toward comprehension abilities and communication difficulties among the older through conversations with their partners. P23 expressed that
\textit{"My previous perception of the older was that their information retrieval might not be very accurate, leading to misunderstandings and biases in their actions, but this did not happen to my partner."} 
Furthermore, six participants shared that the reaction speed and communication skills of their older partners exceeded their initial expectations, even when the performer of the older players slightly slowed down their responses. P12 remarked that \textit{"His responses and communication are both better than what I had in mind. This really surprised me."}

Moreover, the game not only challenged negative stereotypes but also fostered a more positive view of the older through the process of intergenerational interaction. P26 voiced this transformation, stating, \textit{"
He was willing to listen to other people's opinions. Also, he is more knowledgeable than I thought."} Six participants, including P26, emphasized the newfound recognition of the strengths of older individuals, such as their patience and depth of knowledge. 
These findings indicate intergenerational VR collaboration positively influences ageism mitigation by both confronting negative stereotypes and enhancing positive attitudes.

\section{Discussion}
Our research reveals that intergenerational collaboration has a positive effect on both explicit and implicit ageism reduction. 
These findings suggest that intergenerational collaborative games may be a successful intervention to reduce ageism, complementing existing methods such as perspective taking \cite{OH2016398} and virtual embodiment \cite{banakou2018virtually, ayalon2023combatting}.

The VR collaborative gaming has a major effect on ageism-related issues. Oh's research showed how VR interventions can reduce the negative effects of threats against the older \cite{OH2016398}, and our study extends these findings by concluding that VR intergenerational interventions can enhance positive factors such as warmth and competence. Additionally, participants developed more favorable opinions, recognizing the knowledge and patience of the older. 


Qualitative results show the effectiveness of game design in promoting high-quality intergenerational collaboration. The introduction of cognitive and physical activity modules effectively used the strengths of both younger and older players, diminishing social dominance during the collaborative process. 
The participants expressed satisfaction with the collaboration and communication aspects of the game, noting that the collaborative experience was more enjoyable than playing alone. These results support the idea that VR is an effective tool for intergenerational communication and collaboration \cite{baker2019interrogating, wei2023bridging, chou2022empirical}. Additionally, they suggest that the collaborative process can significantly improve the overall gameplay experience.



Our research shows that intergenerational collaboration in a VR environment can effectively reduce ageism. To harness this potential, future VR experiments should consider incorporating older characters and collaborative tasks into their designs. These characters can engage with players and work together with them to complete various tasks within the virtual world. 
By showing the capabilities of older individuals in a collaborative context, designers can challenge stereotypes and foster more inclusive attitudes. Also, our novel game mechanism can be used to decrease social dominance in intergenerational collaboration. By splitting the game into three sections, developers can take advantage of different age groups. 
This design balances the contribution of each age group and encourages a cooperative atmosphere that minimizes social dominance.



The limitation of our study is the absence of actual older participants in the experiment. While our objective was to investigate the reduction of ageism among younger participants, the absence of older individuals in the study is a noteworthy constraint. Our future study will address the limitation by recruiting older participants to gain valuable insights into their perspectives, experiences, and reactions during VR intergenerational collaboration. Their feedback and perceptions are crucial for a more comprehensive understanding of the reduction of ageism in VR contexts and for designing interventions that address the needs and concerns of the older population.


\bibliographystyle{ACM-Reference-Format}
\bibliography{sample-base}


\begin{thebibliography}{48}


\ifx \showCODEN    \undefined \def \showCODEN     #1{\unskip}     \fi
\ifx \showDOI      \undefined \def \showDOI       #1{#1}\fi
\ifx \showISBNx    \undefined \def \showISBNx     #1{\unskip}     \fi
\ifx \showISBNxiii \undefined \def \showISBNxiii  #1{\unskip}     \fi
\ifx \showISSN     \undefined \def \showISSN      #1{\unskip}     \fi
\ifx \showLCCN     \undefined \def \showLCCN      #1{\unskip}     \fi
\ifx \shownote     \undefined \def \shownote      #1{#1}          \fi
\ifx \showarticletitle \undefined \def \showarticletitle #1{#1}   \fi
\ifx \showURL      \undefined \def \showURL       {\relax}        \fi
\providecommand\bibfield[2]{#2}
\providecommand\bibinfo[2]{#2}
\providecommand\natexlab[1]{#1}
\providecommand\showeprint[2][]{arXiv:#2}

\bibitem[Allport et~al\mbox{.}(1954)]%
        {allport1954nature}
\bibfield{author}{\bibinfo{person}{Gordon~Willard Allport},
  \bibinfo{person}{Kenneth Clark}, {and} \bibinfo{person}{Thomas Pettigrew}.}
  \bibinfo{year}{1954}\natexlab{}.
\newblock \showarticletitle{The nature of prejudice}.
\newblock  (\bibinfo{year}{1954}).
\newblock


\bibitem[Ayalon et~al\mbox{.}(2023)]%
        {ayalon2023combatting}
\bibfield{author}{\bibinfo{person}{Liat Ayalon}, \bibinfo{person}{Ehud Dayan},
  {and} \bibinfo{person}{Sara Freedman}.} \bibinfo{year}{2023}\natexlab{}.
\newblock \showarticletitle{Combatting ageism through virtual embodiment? Using
  explicit and implicit measures}.
\newblock \bibinfo{journal}{\emph{International Psychogeriatrics}}
  \bibinfo{volume}{35}, \bibinfo{number}{3} (\bibinfo{year}{2023}),
  \bibinfo{pages}{157--163}.
\newblock


\bibitem[Baker et~al\mbox{.}(2019)]%
        {baker2019interrogating}
\bibfield{author}{\bibinfo{person}{Steven Baker}, \bibinfo{person}{Ryan~M
  Kelly}, \bibinfo{person}{Jenny Waycott}, \bibinfo{person}{Romina Carrasco},
  \bibinfo{person}{Thuong Hoang}, \bibinfo{person}{Frances Batchelor},
  \bibinfo{person}{Elizabeth Ozanne}, \bibinfo{person}{Briony Dow},
  \bibinfo{person}{Jeni Warburton}, {and} \bibinfo{person}{Frank Vetere}.}
  \bibinfo{year}{2019}\natexlab{}.
\newblock \showarticletitle{Interrogating social virtual reality as a
  communication medium for older adults}.
\newblock \bibinfo{journal}{\emph{Proceedings of the ACM on Human-Computer
  Interaction}} \bibinfo{volume}{3}, \bibinfo{number}{CSCW}
  (\bibinfo{year}{2019}), \bibinfo{pages}{1--24}.
\newblock


\bibitem[Baker et~al\mbox{.}(2020)]%
        {baker2020evaluating}
\bibfield{author}{\bibinfo{person}{Steven Baker}, \bibinfo{person}{Jenny
  Waycott}, \bibinfo{person}{Elena Robertson}, \bibinfo{person}{Romina
  Carrasco}, \bibinfo{person}{Barbara~Barbosa Neves}, \bibinfo{person}{Ralph
  Hampson}, {and} \bibinfo{person}{Frank Vetere}.}
  \bibinfo{year}{2020}\natexlab{}.
\newblock \showarticletitle{Evaluating the use of interactive virtual reality
  technology with older adults living in residential aged care}.
\newblock \bibinfo{journal}{\emph{Information Processing \& Management}}
  \bibinfo{volume}{57}, \bibinfo{number}{3} (\bibinfo{year}{2020}),
  \bibinfo{pages}{102105}.
\newblock


\bibitem[Banakou et~al\mbox{.}(2018)]%
        {banakou2018virtually}
\bibfield{author}{\bibinfo{person}{Domna Banakou}, \bibinfo{person}{Sameer
  Kishore}, {and} \bibinfo{person}{Mel Slater}.}
  \bibinfo{year}{2018}\natexlab{}.
\newblock \showarticletitle{Virtually being Einstein results in an improvement
  in cognitive task performance and a decrease in age bias}.
\newblock \bibinfo{journal}{\emph{Frontiers in psychology}}
  \bibinfo{volume}{9} (\bibinfo{year}{2018}), \bibinfo{pages}{917}.
\newblock


\bibitem[Burnes et~al\mbox{.}(2019)]%
        {burnes2019interventions}
\bibfield{author}{\bibinfo{person}{David Burnes}, \bibinfo{person}{Christine
  Sheppard}, \bibinfo{person}{Charles~R Henderson~Jr}, \bibinfo{person}{Monica
  Wassel}, \bibinfo{person}{Richenda Cope}, \bibinfo{person}{Chantal Barber},
  {and} \bibinfo{person}{Karl Pillemer}.} \bibinfo{year}{2019}\natexlab{}.
\newblock \showarticletitle{Interventions to reduce ageism against older
  adults: A systematic review and meta-analysis}.
\newblock \bibinfo{journal}{\emph{American journal of public health}}
  \bibinfo{volume}{109}, \bibinfo{number}{8} (\bibinfo{year}{2019}),
  \bibinfo{pages}{e1--e9}.
\newblock


\bibitem[Cattell(1963)]%
        {cattell1963theory}
\bibfield{author}{\bibinfo{person}{Raymond~B Cattell}.}
  \bibinfo{year}{1963}\natexlab{}.
\newblock \showarticletitle{Theory of fluid and crystallized intelligence: A
  critical experiment.}
\newblock \bibinfo{journal}{\emph{Journal of educational psychology}}
  \bibinfo{volume}{54}, \bibinfo{number}{1} (\bibinfo{year}{1963}),
  \bibinfo{pages}{1}.
\newblock


\bibitem[Chang et~al\mbox{.}(2020)]%
        {chang2020global}
\bibfield{author}{\bibinfo{person}{E-Shien Chang}, \bibinfo{person}{Sneha
  Kannoth}, \bibinfo{person}{Samantha Levy}, \bibinfo{person}{Shi-Yi Wang},
  \bibinfo{person}{John~E Lee}, {and} \bibinfo{person}{Becca~R Levy}.}
  \bibinfo{year}{2020}\natexlab{}.
\newblock \showarticletitle{Global reach of ageism on older persons’ health:
  A systematic review}.
\newblock \bibinfo{journal}{\emph{PloS one}} \bibinfo{volume}{15},
  \bibinfo{number}{1} (\bibinfo{year}{2020}), \bibinfo{pages}{e0220857}.
\newblock


\bibitem[Chou et~al\mbox{.}(2022)]%
        {chou2022empirical}
\bibfield{author}{\bibinfo{person}{Wen-Huei Chou}, \bibinfo{person}{Yi-Chun
  Li}, \bibinfo{person}{Ya-Fang Chen}, \bibinfo{person}{Mieko Ohsuga}, {and}
  \bibinfo{person}{Tsuyoshi Inoue}.} \bibinfo{year}{2022}\natexlab{}.
\newblock \showarticletitle{Empirical study of virtual reality to promote
  intergenerational communication: Taiwan traditional glove puppetry as
  example}.
\newblock \bibinfo{journal}{\emph{Sustainability}} \bibinfo{volume}{14},
  \bibinfo{number}{6} (\bibinfo{year}{2022}), \bibinfo{pages}{3213}.
\newblock


\bibitem[Coyle et~al\mbox{.}(2015)]%
        {coyle2015computerized}
\bibfield{author}{\bibinfo{person}{Hannah Coyle}, \bibinfo{person}{Victoria
  Traynor}, {and} \bibinfo{person}{Nadia Solowij}.}
  \bibinfo{year}{2015}\natexlab{}.
\newblock \showarticletitle{Computerized and virtual reality cognitive training
  for individuals at high risk of cognitive decline: systematic review of the
  literature}.
\newblock \bibinfo{journal}{\emph{The American Journal of Geriatric
  Psychiatry}} \bibinfo{volume}{23}, \bibinfo{number}{4}
  (\bibinfo{year}{2015}), \bibinfo{pages}{335--359}.
\newblock


\bibitem[Dooley and Frankel(1990)]%
        {dooley1990improving}
\bibfield{author}{\bibinfo{person}{Stephen Dooley} {and}
  \bibinfo{person}{B~Gail Frankel}.} \bibinfo{year}{1990}\natexlab{}.
\newblock \showarticletitle{Improving attitudes toward elderly people:
  Evaluation of an intervention program for adolescents}.
\newblock \bibinfo{journal}{\emph{Canadian Journal on Aging/La Revue canadienne
  du vieillissement}} \bibinfo{volume}{9}, \bibinfo{number}{4}
  (\bibinfo{year}{1990}), \bibinfo{pages}{400--409}.
\newblock


\bibitem[Fiske et~al\mbox{.}(2018)]%
        {fiske2018model}
\bibfield{author}{\bibinfo{person}{Susan~T Fiske}, \bibinfo{person}{Amy~JC
  Cuddy}, \bibinfo{person}{Peter Glick}, {and} \bibinfo{person}{Jun Xu}.}
  \bibinfo{year}{2018}\natexlab{}.
\newblock \showarticletitle{A model of (often mixed) stereotype content:
  Competence and warmth respectively follow from perceived status and
  competition}.
\newblock In \bibinfo{booktitle}{\emph{Social cognition}}.
  \bibinfo{publisher}{Routledge}, \bibinfo{pages}{162--214}.
\newblock


\bibitem[Fiske et~al\mbox{.}(1999)]%
        {fiske1999dis}
\bibfield{author}{\bibinfo{person}{Susan~T Fiske}, \bibinfo{person}{Juan Xu},
  \bibinfo{person}{Amy~C Cuddy}, {and} \bibinfo{person}{Peter Glick}.}
  \bibinfo{year}{1999}\natexlab{}.
\newblock \showarticletitle{(Dis) respecting versus (dis) liking: Status and
  interdependence predict ambivalent stereotypes of competence and warmth}.
\newblock \bibinfo{journal}{\emph{Journal of social issues}}
  \bibinfo{volume}{55}, \bibinfo{number}{3} (\bibinfo{year}{1999}),
  \bibinfo{pages}{473--489}.
\newblock


\bibitem[Greenwald et~al\mbox{.}(1998)]%
        {greenwald1998measuring}
\bibfield{author}{\bibinfo{person}{Anthony~G Greenwald},
  \bibinfo{person}{Debbie~E McGhee}, {and} \bibinfo{person}{Jordan~LK
  Schwartz}.} \bibinfo{year}{1998}\natexlab{}.
\newblock \showarticletitle{Measuring individual differences in implicit
  cognition: the implicit association test.}
\newblock \bibinfo{journal}{\emph{Journal of personality and social
  psychology}} \bibinfo{volume}{74}, \bibinfo{number}{6}
  (\bibinfo{year}{1998}), \bibinfo{pages}{1464}.
\newblock


\bibitem[Hausknecht et~al\mbox{.}(2017)]%
        {hausknecht2017blurring}
\bibfield{author}{\bibinfo{person}{Simone Hausknecht}, \bibinfo{person}{Carman
  Neustaedter}, {and} \bibinfo{person}{David Kaufman}.}
  \bibinfo{year}{2017}\natexlab{}.
\newblock \showarticletitle{Blurring the lines of age: intergenerational
  collaboration in alternate reality games}.
\newblock \bibinfo{journal}{\emph{Game-based learning across the lifespan:
  Cross-generational and age-oriented topics}} (\bibinfo{year}{2017}),
  \bibinfo{pages}{47--64}.
\newblock


\bibitem[Hill et~al\mbox{.}(2017)]%
        {hill2017computerized}
\bibfield{author}{\bibinfo{person}{Nicole~TM Hill}, \bibinfo{person}{Loren
  Mowszowski}, \bibinfo{person}{Sharon~L Naismith}, \bibinfo{person}{Verity~L
  Chadwick}, \bibinfo{person}{Michael Valenzuela}, {and} \bibinfo{person}{Amit
  Lampit}.} \bibinfo{year}{2017}\natexlab{}.
\newblock \showarticletitle{Computerized cognitive training in older adults
  with mild cognitive impairment or dementia: a systematic review and
  meta-analysis}.
\newblock \bibinfo{journal}{\emph{American Journal of Psychiatry}}
  \bibinfo{volume}{174}, \bibinfo{number}{4} (\bibinfo{year}{2017}),
  \bibinfo{pages}{329--340}.
\newblock


\bibitem[Hodgkins(1962)]%
        {hodgkins1962influence}
\bibfield{author}{\bibinfo{person}{Jean Hodgkins}.}
  \bibinfo{year}{1962}\natexlab{}.
\newblock \showarticletitle{Influence of age on the speed of reaction and
  movement in females.}
\newblock \bibinfo{journal}{\emph{Journal of Gerontology}}
  (\bibinfo{year}{1962}).
\newblock


\bibitem[H{\o}eg et~al\mbox{.}(2023)]%
        {hoeg2023buddy}
\bibfield{author}{\bibinfo{person}{Emil~Rosenlund H{\o}eg},
  \bibinfo{person}{Jon~Ram Bruun-Pedersen}, \bibinfo{person}{Shannon Cheary},
  \bibinfo{person}{Lars~Koreska Andersen}, \bibinfo{person}{Razvan Paisa},
  \bibinfo{person}{Stefania Serafin}, {and} \bibinfo{person}{Belinda Lange}.}
  \bibinfo{year}{2023}\natexlab{}.
\newblock \showarticletitle{Buddy biking: a user study on social collaboration
  in a virtual reality exergame for rehabilitation}.
\newblock \bibinfo{journal}{\emph{Virtual Reality}} \bibinfo{volume}{27},
  \bibinfo{number}{1} (\bibinfo{year}{2023}), \bibinfo{pages}{245--262}.
\newblock


\bibitem[Hultsch et~al\mbox{.}(2002)]%
        {hultsch2002variability}
\bibfield{author}{\bibinfo{person}{David~F Hultsch}, \bibinfo{person}{Stuart~WS
  MacDonald}, {and} \bibinfo{person}{Roger~A Dixon}.}
  \bibinfo{year}{2002}\natexlab{}.
\newblock \showarticletitle{Variability in reaction time performance of younger
  and older adults}.
\newblock \bibinfo{journal}{\emph{The Journals of Gerontology Series B:
  Psychological Sciences and Social Sciences}} \bibinfo{volume}{57},
  \bibinfo{number}{2} (\bibinfo{year}{2002}), \bibinfo{pages}{P101--P115}.
\newblock


\bibitem[Jarrott et~al\mbox{.}(2022)]%
        {jarrott2022intergenerational}
\bibfield{author}{\bibinfo{person}{Shannon~E Jarrott}, \bibinfo{person}{Skye~N
  Leedahl}, \bibinfo{person}{Tamar~E Shovali}, \bibinfo{person}{Carson
  De~Fries}, \bibinfo{person}{Amy DelPo}, \bibinfo{person}{Erica Estus},
  \bibinfo{person}{Caroline Gangji}, \bibinfo{person}{Leslie Hasche},
  \bibinfo{person}{Jill Juris}, \bibinfo{person}{Roddy MacInnes},
  {et~al\mbox{.}}} \bibinfo{year}{2022}\natexlab{}.
\newblock \showarticletitle{Intergenerational programming during the pandemic:
  Transformation during (constantly) changing times}.
\newblock \bibinfo{journal}{\emph{Journal of Social Issues}}
  \bibinfo{volume}{78}, \bibinfo{number}{4} (\bibinfo{year}{2022}),
  \bibinfo{pages}{1038--1065}.
\newblock


\bibitem[Jerald(2015)]%
        {jerald2015vr}
\bibfield{author}{\bibinfo{person}{Jason Jerald}.}
  \bibinfo{year}{2015}\natexlab{}.
\newblock \bibinfo{booktitle}{\emph{The VR book: Human-centered design for
  virtual reality}}.
\newblock \bibinfo{publisher}{Morgan \& Claypool}.
\newblock


\bibitem[Khoo et~al\mbox{.}(2008)]%
        {khoo2008age}
\bibfield{author}{\bibinfo{person}{Eng~Tat Khoo}, \bibinfo{person}{Adrian~David
  Cheok}, \bibinfo{person}{Ta~Huynh~Duy Nguyen}, {and} \bibinfo{person}{Zhigeng
  Pan}.} \bibinfo{year}{2008}\natexlab{}.
\newblock \showarticletitle{Age invaders: social and physical
  inter-generational mixed reality family entertainment}.
\newblock \bibinfo{journal}{\emph{Virtual Reality}}  \bibinfo{volume}{12}
  (\bibinfo{year}{2008}), \bibinfo{pages}{3--16}.
\newblock


\bibitem[Khoo et~al\mbox{.}(2010)]%
        {khoo2010designing}
\bibfield{author}{\bibinfo{person}{Eng~Tat Khoo}, \bibinfo{person}{Tim
  Merritt}, {and} \bibinfo{person}{Adrian~David Cheok}.}
  \bibinfo{year}{2010}\natexlab{}.
\newblock \showarticletitle{Designing a mixed reality intergenerational
  entertainment system}.
\newblock \bibinfo{journal}{\emph{The Engineering of Mixed Reality Systems}}
  (\bibinfo{year}{2010}), \bibinfo{pages}{121--141}.
\newblock


\bibitem[Kogan(1961)]%
        {kogan1961attitudes}
\bibfield{author}{\bibinfo{person}{Nathan Kogan}.}
  \bibinfo{year}{1961}\natexlab{}.
\newblock \showarticletitle{Attitudes toward old people: the development of a
  scale and an examination of correlates.}
\newblock \bibinfo{journal}{\emph{The Journal of Abnormal and Social
  Psychology}} \bibinfo{volume}{62}, \bibinfo{number}{1}
  (\bibinfo{year}{1961}), \bibinfo{pages}{44}.
\newblock


\bibitem[Levy and Apriceno(2019)]%
        {levy2019ageing}
\bibfield{author}{\bibinfo{person}{Sheri Levy} {and} \bibinfo{person}{MaryBeth
  Apriceno}.} \bibinfo{year}{2019}\natexlab{}.
\newblock \showarticletitle{Ageing: The role of ageism}.
\newblock \bibinfo{journal}{\emph{OBM Geriatrics}} \bibinfo{volume}{3},
  \bibinfo{number}{4} (\bibinfo{year}{2019}), \bibinfo{pages}{1--16}.
\newblock


\bibitem[Loe(2013)]%
        {loe2013digital}
\bibfield{author}{\bibinfo{person}{Meika Loe}.}
  \bibinfo{year}{2013}\natexlab{}.
\newblock \showarticletitle{The digital life history project: Intergenerational
  collaborative research}.
\newblock \bibinfo{journal}{\emph{Gerontology \& Geriatrics Education}}
  \bibinfo{volume}{34}, \bibinfo{number}{1} (\bibinfo{year}{2013}),
  \bibinfo{pages}{26--42}.
\newblock


\bibitem[Murphy-Russell et~al\mbox{.}(1986)]%
        {murphy1986changing}
\bibfield{author}{\bibinfo{person}{Sheila Murphy-Russell},
  \bibinfo{person}{Ann~H Die}, {and} \bibinfo{person}{James~L Walker~Jr}.}
  \bibinfo{year}{1986}\natexlab{}.
\newblock \showarticletitle{Changing attitudes toward the elderly: The impact
  of three methods of attitude change}.
\newblock \bibinfo{journal}{\emph{Educational Gerontology}}
  \bibinfo{volume}{12}, \bibinfo{number}{3} (\bibinfo{year}{1986}),
  \bibinfo{pages}{241--251}.
\newblock


\bibitem[Nelson(2010)]%
        {nelson2010ageism}
\bibfield{author}{\bibinfo{person}{Todd~D Nelson}.}
  \bibinfo{year}{2010}\natexlab{}.
\newblock \showarticletitle{Ageism: The strange case of prejudice against the
  older you}.
\newblock In \bibinfo{booktitle}{\emph{Disability and aging discrimination:
  Perspectives in law and psychology}}. \bibinfo{publisher}{Springer},
  \bibinfo{pages}{37--47}.
\newblock


\bibitem[Nosek et~al\mbox{.}(2005)]%
        {nosek2005understanding}
\bibfield{author}{\bibinfo{person}{Brian~A Nosek}, \bibinfo{person}{Anthony~G
  Greenwald}, {and} \bibinfo{person}{Mahzarin~R Banaji}.}
  \bibinfo{year}{2005}\natexlab{}.
\newblock \showarticletitle{Understanding and using the Implicit Association
  Test: II. Method variables and construct validity}.
\newblock \bibinfo{journal}{\emph{Personality and Social Psychology Bulletin}}
  \bibinfo{volume}{31}, \bibinfo{number}{2} (\bibinfo{year}{2005}),
  \bibinfo{pages}{166--180}.
\newblock


\bibitem[Oh et~al\mbox{.}(2016)]%
        {OH2016398}
\bibfield{author}{\bibinfo{person}{Soo~Youn Oh}, \bibinfo{person}{Jeremy
  Bailenson}, \bibinfo{person}{Erika Weisz}, {and} \bibinfo{person}{Jamil
  Zaki}.} \bibinfo{year}{2016}\natexlab{}.
\newblock \showarticletitle{Virtually old: Embodied perspective taking and the
  reduction of ageism under threat}.
\newblock \bibinfo{journal}{\emph{Computers in Human Behavior}}
  \bibinfo{volume}{60} (\bibinfo{year}{2016}), \bibinfo{pages}{398--410}.
\newblock
\showISSN{0747-5632}
\urldef\tempurl%
\url{https://doi.org/10.1016/j.chb.2016.02.007}
\showDOI{\tempurl}


\bibitem[Organization et~al\mbox{.}(2016)]%
        {world2016multisectoral}
\bibfield{author}{\bibinfo{person}{World~Health Organization} {et~al\mbox{.}}}
  \bibinfo{year}{2016}\natexlab{}.
\newblock \showarticletitle{Multisectoral action for a life course approach to
  healthy ageing: draft global strategy and plan of action on ageing and
  health}.
\newblock \bibinfo{journal}{\emph{World Health Organization: Geneva,
  Switzerland}} (\bibinfo{year}{2016}), \bibinfo{pages}{1--37}.
\newblock


\bibitem[Pierson and Montoye(1958)]%
        {pierson1958movement}
\bibfield{author}{\bibinfo{person}{William~R Pierson} {and}
  \bibinfo{person}{Henry~J Montoye}.} \bibinfo{year}{1958}\natexlab{}.
\newblock \showarticletitle{Movement time, reaction time, and age.}
\newblock \bibinfo{journal}{\emph{Journal of Gerontology}}
  (\bibinfo{year}{1958}).
\newblock


\bibitem[Romero and Lille(2017)]%
        {romero2017intergenerational}
\bibfield{author}{\bibinfo{person}{Margarida Romero} {and}
  \bibinfo{person}{Benjamin Lille}.} \bibinfo{year}{2017}\natexlab{}.
\newblock \showarticletitle{Intergenerational techno-creative activities in a
  library fablab}. In \bibinfo{booktitle}{\emph{Human Aspects of IT for the
  Aged Population. Applications, Services and Contexts: Third International
  Conference, ITAP 2017, Held as Part of HCI International 2017, Vancouver, BC,
  Canada, July 9-14, 2017, Proceedings, Part II 3}}. Springer,
  \bibinfo{pages}{526--536}.
\newblock


\bibitem[Rubin et~al\mbox{.}(2015)]%
        {rubin2015}
\bibfield{author}{\bibinfo{person}{Sadie~E Rubin}, \bibinfo{person}{Tracey~L
  Gendron}, \bibinfo{person}{Cortney~A Wren}, \bibinfo{person}{Kelechi~C
  Ogbonna}, \bibinfo{person}{Ernest~G Gonzales}, {and} \bibinfo{person}{Emily~P
  Peron}.} \bibinfo{year}{2015}\natexlab{}.
\newblock \showarticletitle{Challenging gerontophobia and ageism through a
  collaborative intergenerational art program}.
\newblock \bibinfo{journal}{\emph{Journal of Intergenerational Relationships}}
  \bibinfo{volume}{13}, \bibinfo{number}{3} (\bibinfo{year}{2015}),
  \bibinfo{pages}{241--254}.
\newblock


\bibitem[Salthouse(2012)]%
        {salthouse2012consequences}
\bibfield{author}{\bibinfo{person}{Timothy Salthouse}.}
  \bibinfo{year}{2012}\natexlab{}.
\newblock \showarticletitle{Consequences of age-related cognitive declines}.
\newblock \bibinfo{journal}{\emph{Annual review of psychology}}
  \bibinfo{volume}{63} (\bibinfo{year}{2012}), \bibinfo{pages}{201--226}.
\newblock


\bibitem[Saredakis et~al\mbox{.}(2020)]%
        {saredakis2020using}
\bibfield{author}{\bibinfo{person}{Dimitrios Saredakis},
  \bibinfo{person}{Hannah~AD Keage}, \bibinfo{person}{Megan Corlis}, {and}
  \bibinfo{person}{Tobias Loetscher}.} \bibinfo{year}{2020}\natexlab{}.
\newblock \showarticletitle{Using virtual reality to improve apathy in
  residential aged care: Mixed methods study}.
\newblock \bibinfo{journal}{\emph{Journal of medical Internet research}}
  \bibinfo{volume}{22}, \bibinfo{number}{6} (\bibinfo{year}{2020}),
  \bibinfo{pages}{e17632}.
\newblock


\bibitem[Sherman and Craig(2018)]%
        {sherman2018understanding}
\bibfield{author}{\bibinfo{person}{William~R Sherman} {and}
  \bibinfo{person}{Alan~B Craig}.} \bibinfo{year}{2018}\natexlab{}.
\newblock \bibinfo{booktitle}{\emph{Understanding virtual reality: Interface,
  application, and design}}.
\newblock \bibinfo{publisher}{Morgan Kaufmann}.
\newblock


\bibitem[Springate et~al\mbox{.}(2008)]%
        {springate2008intergenerational}
\bibfield{author}{\bibinfo{person}{Iain Springate}, \bibinfo{person}{Mary
  Atkinson}, {and} \bibinfo{person}{Kerry Martin}.}
  \bibinfo{year}{2008}\natexlab{}.
\newblock \bibinfo{booktitle}{\emph{Intergenerational Practice: A Review of the
  Literature. LGA Research Report F/SR262.}}
\newblock \bibinfo{publisher}{ERIC}.
\newblock


\bibitem[Stephan and Stephan(2013)]%
        {stephan2013integrated}
\bibfield{author}{\bibinfo{person}{Walter~G Stephan} {and}
  \bibinfo{person}{Cookie~White Stephan}.} \bibinfo{year}{2013}\natexlab{}.
\newblock \showarticletitle{An integrated threat theory of prejudice}.
\newblock In \bibinfo{booktitle}{\emph{Reducing prejudice and discrimination}}.
  \bibinfo{publisher}{Psychology Press}, \bibinfo{pages}{23--45}.
\newblock


\bibitem[Stephan et~al\mbox{.}(1999)]%
        {stephan1999anxiety}
\bibfield{author}{\bibinfo{person}{Walter~G Stephan},
  \bibinfo{person}{Cookie~White Stephan}, {and} \bibinfo{person}{William~B
  Gudykunst}.} \bibinfo{year}{1999}\natexlab{}.
\newblock \showarticletitle{Anxiety in intergroup relations: A comparison of
  anxiety/uncertainty management theory and integrated threat theory}.
\newblock \bibinfo{journal}{\emph{International Journal of Intercultural
  Relations}} \bibinfo{volume}{23}, \bibinfo{number}{4} (\bibinfo{year}{1999}),
  \bibinfo{pages}{613--628}.
\newblock


\bibitem[Tassinari et~al\mbox{.}(2022)]%
        {tassinari2022use}
\bibfield{author}{\bibinfo{person}{Matilde Tassinari},
  \bibinfo{person}{Matthias~Burkard Aulbach}, {and} \bibinfo{person}{Inga
  Jasinskaja-Lahti}.} \bibinfo{year}{2022}\natexlab{}.
\newblock \showarticletitle{The use of virtual reality in studying prejudice
  and its reduction: A systematic review}.
\newblock \bibinfo{journal}{\emph{PloS one}} \bibinfo{volume}{17},
  \bibinfo{number}{7} (\bibinfo{year}{2022}), \bibinfo{pages}{e0270748}.
\newblock


\bibitem[Tausch et~al\mbox{.}(2007)]%
        {tausch2007individual}
\bibfield{author}{\bibinfo{person}{Nicole Tausch}, \bibinfo{person}{Tania Tam},
  \bibinfo{person}{Miles Hewstone}, \bibinfo{person}{Jared Kenworthy}, {and}
  \bibinfo{person}{Ed Cairns}.} \bibinfo{year}{2007}\natexlab{}.
\newblock \showarticletitle{Individual-level and group-level mediators of
  contact effects in Northern Ireland: The moderating role of social
  identification}.
\newblock \bibinfo{journal}{\emph{British journal of social psychology}}
  \bibinfo{volume}{46}, \bibinfo{number}{3} (\bibinfo{year}{2007}),
  \bibinfo{pages}{541--556}.
\newblock


\bibitem[United(2018)]%
        {united2018all}
\bibfield{author}{\bibinfo{person}{Generations United}.}
  \bibinfo{year}{2018}\natexlab{}.
\newblock \showarticletitle{All in together: Creating places where young and
  old thrive}.
\newblock \bibinfo{journal}{\emph{Generations United: Washington, DC, USA}}
  (\bibinfo{year}{2018}).
\newblock


\bibitem[Van Houwelingen-Snippe et~al\mbox{.}(2021)]%
        {van2021virtual}
\bibfield{author}{\bibinfo{person}{Josca Van Houwelingen-Snippe},
  \bibinfo{person}{Somaya Ben~Allouch}, {and} \bibinfo{person}{Thomas~JL
  Van~Rompay}.} \bibinfo{year}{2021}\natexlab{}.
\newblock \showarticletitle{Virtual reality representations of nature to
  improve well-being amongst older adults: a rapid review}.
\newblock \bibinfo{journal}{\emph{Journal of Technology in Behavioral Science}}
   \bibinfo{volume}{6} (\bibinfo{year}{2021}), \bibinfo{pages}{464--485}.
\newblock


\bibitem[Wang et~al\mbox{.}(2022)]%
        {su14106067}
\bibfield{author}{\bibinfo{person}{Chao-Ming Wang}, \bibinfo{person}{Cheng-Hao
  Shao}, {and} \bibinfo{person}{Cheng-En Han}.}
  \bibinfo{year}{2022}\natexlab{}.
\newblock \showarticletitle{Construction of a Tangible VR-Based Interactive
  System for Intergenerational Learning}.
\newblock \bibinfo{journal}{\emph{Sustainability}} \bibinfo{volume}{14},
  \bibinfo{number}{10} (\bibinfo{year}{2022}).
\newblock
\showISSN{2071-1050}
\urldef\tempurl%
\url{https://doi.org/10.3390/su14106067}
\showDOI{\tempurl}


\bibitem[Waycott et~al\mbox{.}(2022)]%
        {waycott2022role}
\bibfield{author}{\bibinfo{person}{Jenny Waycott}, \bibinfo{person}{Ryan~M
  Kelly}, \bibinfo{person}{Steven Baker}, \bibinfo{person}{Barbara
  Barbosa~Neves}, \bibinfo{person}{Kong~Saoane Thach}, {and}
  \bibinfo{person}{Reeva Lederman}.} \bibinfo{year}{2022}\natexlab{}.
\newblock \showarticletitle{The role of staff in facilitating immersive virtual
  reality for enrichment in aged care: an ethic of care perspective}. In
  \bibinfo{booktitle}{\emph{Proceedings of the 2022 CHI Conference on Human
  Factors in Computing Systems}}. \bibinfo{pages}{1--17}.
\newblock


\bibitem[Wei et~al\mbox{.}(2023)]%
        {wei2023bridging}
\bibfield{author}{\bibinfo{person}{Xiaoying Wei}, \bibinfo{person}{Yizheng Gu},
  \bibinfo{person}{Emily Kuang}, \bibinfo{person}{Xian Wang},
  \bibinfo{person}{Beiyan Cao}, \bibinfo{person}{Xiaofu Jin}, {and}
  \bibinfo{person}{Mingming Fan}.} \bibinfo{year}{2023}\natexlab{}.
\newblock \showarticletitle{Bridging the Generational Gap: Exploring How
  Virtual Reality Supports Remote Communication Between Grandparents and
  Grandchildren}. In \bibinfo{booktitle}{\emph{Proceedings of the 2023 CHI
  Conference on Human Factors in Computing Systems}}. \bibinfo{pages}{1--15}.
\newblock


\bibitem[Xu et~al\mbox{.}(2023)]%
        {xu2023designing}
\bibfield{author}{\bibinfo{person}{Tong~Bill Xu}, \bibinfo{person}{Armin
  Mostafavi}, \bibinfo{person}{Benjamin~C Kim}, \bibinfo{person}{Angella~Anyi
  Lee}, \bibinfo{person}{Walter Boot}, \bibinfo{person}{Sara Czaja}, {and}
  \bibinfo{person}{Saleh Kalantari}.} \bibinfo{year}{2023}\natexlab{}.
\newblock \showarticletitle{Designing Virtual Environments for Social
  Engagement in Older Adults: A Qualitative Multi-site Study}. In
  \bibinfo{booktitle}{\emph{Proceedings of the 2023 CHI Conference on Human
  Factors in Computing Systems}}. \bibinfo{pages}{1--15}.
\newblock


\end{thebibliography}

\end{document}